\documentstyle{mn}           

\newif\ifAMStwofonts
\AMStwofontstrue

\newcommand{\mic}{\hbox{\,$\mu$m}}
\newcommand{\iron}{\hbox{[Fe\,{\sc ii}]}}
\newcommand{\ironsix}{\iron\ 1.644\mic}
\newcommand{\irontwo}{\iron\ 1.257\mic}

\newcommand{\brg}{\hbox{Br${\gamma}$}}                

\newcommand{\hh}{\hbox{H$_2$}}                       

\newcommand{\kms}{\hbox{\,km\,s$^{-1}$}}                
\newcommand{\HI}{\hbox{H\,{\sc i}}}                   
\newcommand{\cc}{\hbox{cm$^{-3}$}}                   
\newcommand{\cmm}{\hbox{cm$^{-2}$}}

\newcommand{\ergs}{\hbox{erg\,s$^{-1}$\,\cmm\,sr$^{-1}$}}

\newcommand{\p}{\mbox{$\pm$}}                     

\ifoldfss
  \ifCUPmtlplainloaded \else
    \NewTextAlphabet{textbfit} {cmbxti10} {}
    \NewTextAlphabet{textbfss} {cmssbx10} {}
    \NewMathAlphabet{mathbfit} {cmbxti10} {} 
    \NewMathAlphabet{mathbfss} {cmssbx10} {} 
  \fi
  \ifAMStwofonts
    \ifCUPmtlplainloaded \else
      \NewSymbolFont{upmath} {eurm10}
      \NewSymbolFont{AMSa} {msam10}
      \NewMathSymbol{\upi}     {0}{upmath}{19}
      \NewMathSymbol{\umu}     {0}{upmath}{16}
      \NewMathSymbol{\upartial}{0}{upmath}{40}
      \NewMathSymbol{\leqslant}{3}{AMSa}{36}
      \NewMathSymbol{\geqslant}{3}{AMSa}{3E}

    \fi
  \fi
\fi 

\ifnfssone
  \newmathalphabet{\mathit}
  \addtoversion{normal}{\mathit}{cmr}{m}{it}
  \addtoversion{bold}{\mathit}{cmr}{bx}{it}
  \newmathalphabet{\mathbfit} 
  \addtoversion{normal}{\mathbfit}{cmr}{bx}{it}
  \addtoversion{bold}{\mathbfit}{cmr}{bx}{it}
  \newmathalphabet{\mathbfss} 
  \addtoversion{normal}{\mathbfss}{cmss}{bx}{n}
  \addtoversion{bold}{\mathbfss}{cmss}{bx}{n}
  \ifAMStwofonts
    \ifCUPmtlplainloaded \else
      \UseAMStwoboldmath
      \makeatletter
      \new@mathgroup\upmath@group
      \define@mathgroup\mv@normal\upmath@group{eur}{m}{n}
      \define@mathgroup\mv@bold\upmath@group{eur}{b}{n}
      \edef\UPM{\hexnumber\upmath@group}
      \new@mathgroup\amsa@group
      \define@mathgroup\mv@normal\amsa@group{msa}{m}{n}
      \define@mathgroup\mv@bold\amsa@group{msa}{m}{n}
      \edef\AMSa{\hexnumber\amsa@group}
      \makeatother
      \mathchardef\upi="0\UPM19
      \mathchardef\umu="0\UPM16
      \mathchardef\upartial="0\UPM40
      \mathchardef\leqslant="3\AMSa36
      \mathchardef\geqslant="3\AMSa3E
    \fi
  \fi
\fi 

\ifnfsstwo
  \DeclareMathAlphabet{\mathbfit}{OT1}{cmr}{bx}{it}
  \SetMathAlphabet\mathbfit{bold}{OT1}{cmr}{bx}{it}
  \DeclareMathAlphabet{\mathbfss}{OT1}{cmss}{bx}{n}
  \SetMathAlphabet\mathbfss{bold}{OT1}{cmss}{bx}{n}
  \ifAMStwofonts
    \ifCUPmtlplainloaded \else
      \DeclareSymbolFont{UPM}{U}{eur}{m}{n}
      \SetSymbolFont{UPM}{bold}{U}{eur}{b}{n}
      \DeclareSymbolFont{AMSa}{U}{msa}{m}{n}
      \DeclareMathSymbol{\upi}{0}{UPM}{"19}
      \DeclareMathSymbol{\umu}{0}{UPM}{"16}
      \DeclareMathSymbol{\upartial}{0}{UPM}{"40}
      \DeclareMathSymbol{\leqslant}{3}{AMSa}{"36}
      \DeclareMathSymbol{\geqslant}{3}{AMSa}{"3E}
    \fi
  \fi
\fi 

\ifCUPmtlplainloaded \else
  \ifAMStwofonts \else 
    \def\upi{\pi}
    \def\umu{\mu}
    \def\upartial{\partial}
  \fi
\fi

\title[Excitation and kinematics in the HH\,46/47 bipolar outflow]
{The excitation and kinematical properties of \hh\ and \iron\ in the
HH\,46/47 bipolar outflow\thanks{Based on observations collected at
the European Southern Observatory, La Silla, Chile}}

\author[A.J.L. Fernandes] {A.J.L. Fernandes\\
E-mail: amadeu@astro.up.pt\\
Centro de Astrof\'{\i}sica da Universidade do Porto, Rua das Estrelas, 
4150-762 Porto, Portugal\\
Instituto Superior da Maia, Portugal
}

\begin{document}

\maketitle

\begin{abstract}
  
Long slit spectra of the molecular outflow Herbig-Haro (HH) 46/47
has been taken in the J and K near-infrared bands. The observed \hh\ 
line emission confirms the existence of a bright and extended
redshifted counter-jet outflow southwest of HH46.  We show that this
outflow seems to be composed of two different emission regions which
have distinct heliocentric velocities, in contrast with the optical
appearance, and which implies an acceleration of the counter-jet.  

The observed \iron\ emission suggests an average extinction of 7--9 visual 
magnitudes for the region associated  with the counter-jet.

Through position-velocity diagrams, we show the existence of different
morphologies for the \hh\ and \iron\ emission regions in the
northern part of the HH\,46/47 outflow.  We have detected for the
first time high velocity (-250\kms) \iron\ emission in the region
bridging HH46 to HH47A. The two strong peaks detected can be
identified with the optical positions B8 and HH47B.
  
The \hh\ excitation diagrams for the counter-jet shock suggest an
excitation temperature for the gas of $T_{\rm{ex}} \approx 2600$.
The lack of emission from the higher energy \hh\ lines such as the
4-3~S(3) transition, suggests a thermal excitation scenario for the
origin of the observed emission. Comparison of the \hh\ line ratios
to various shock models yielded useful constraints
about the geometry and type of these shocks. Planar shocks can be
ruled out whereas curved or bow-shocks (both J and C-type) can be
parametrised to adjust our data.

\end{abstract}


\section{Introduction}

For its tight spatial alignment of the emission knots and its clear
visibility against the progenitor dark globule, HH\,46/47 is one of
the most remarkable examples of the classical Herbig-Haro object.
Being optically bright, this HH complex has been the subject of many
studies of its spectral, photometric and kinematical properties after
its discovery by Schwartz (1977).

The morphology of the Gum nebula region (Dopita et~al. 1982) shows the
existence of a very collimated jet that is composed of several
components which extend about 0.39 pc on the sky at the assumed
distance of 350 pc (Eisl{\"o}ffel and Mundt 1994). A young T Tauri
star still obscured by the surrounding dust envelope sits at the
center of this system.  This star illuminates a cloud of dust
revealing a reflection nebula from which the optically visible HH\,46
and HH\,47(A,B,C,D) jet components emerge.

The bipolar nature of this outflow was first ascertained by Graham and
Elias (1983), on discovering blue and redshifted outflowing shocked
gas to either side of the central source.  Schwartz et al.  (1984)
reported the first proper motions for some dense knots along the flow
suggesting the occurrence of episodic ejections of gas material from
the energy source. Combining the radial velocity measurements (Meaburn
\& Dyson 1987, Hartigan et al. 1993, Morse et al 1994) with a detailed
study of the proper motions of the knots in the HH\,46/47 jet
(Eisl{\"o}ffel \& Mundt 1994), a space velocity of about 300\kms is
found for the jet.

While several imaging, spectroscopic and proper motion studies have
been carried out in the optical for the blueshifted part of HH\,46/47,
there has been considerably less work done for the redshifted part of
the flow.  This well-collimated counter-jet flow located southwest of
the source was discovered by Reipurth and Heathcote (1991) in [SII]
$\lambda\lambda$6717/6731 line emission and found to be redshifted
with a heliocentric radial velocity of about +125\kms.
K-band imaging of the redshifted counter-flow has been performed by
Eisl\"offel et al.  (1994), together with \hh\ 1-0~S(1) (2.1218\mic)
and \iron\ (1.644\mic) line imaging. More recently, proper motion
studies of the \hh\ emission have been carried out for the \hh\ knots
in HH\,46/47 (Micono et~al. 1998).

\section{Observations and data reduction}
\label{obs}

The HH\,46/47 complex was observed in 1995 March 14 to 17 using the
cryogenically cooled grating spectrometer IRSPEC installed on the ESO
3.5\,m New Technology Telescope (NTT) sited at La Silla, Chile.  The
instrument was equipped with an InSb CCD detector comprising $58\times
62$ square pixels which provided a pixel scale of 2.2 arcseconds on
the sky.  We have positioned the 2 arcminute long $\times$ 2.2
arcseconds wide slit at a position angle of 57 degrees over the
HH\,46/47 bipolar flow, centered on approximately the HH\,46 emission
nebula continuum (Figure~1). 
This orientation on the sky coincides with the outflow axis of both
optical and infrared maps of HH\,46/47 and the 2 arcminute long slit
allowed us to measure simultaneously the emission region from the
HH\,47A object to the counter-jet flow.  The precise positioning of
the observed slit position was set by peaking up the \hh\ emission
maximum.  The telescope pointing accuracy, normally better than
1\arcsec\ (Moorwood et~al. 1991) has been checked with observations of
the standard stars.

The seeing disk remained under 1\arcsec\ with an average FWHM of
0.8\arcsec.  Grating no.  2 was used with the slit set to a width of 2
pixels.  The spectral resolution was $\sim 0.001$\mic\ (FWHM) at
2.122\mic\ but varied little over the whole K band region ($R =
\lambda / \Delta\lambda \approx 2300$).  For the two observed \iron\ 
transitions, the resolution was $R \approx 3300$ and 1600 at
1.257\mic\ and 1.644\mic, respectively.  At these resolutions, every
frame contained a very short wavelength range and only one emission
line was registered per frame.  Because of the extent of the HH\,46/47
outflow on the 2\arcmin\ slit, explicit sky observations were taken by
offsetting the telescope 40\arcsec\ south from the nominal object
position.  The integration times were 60 seconds on source and 60
seconds on sky and a final frame was determined by averaging many
integrations.  The observational procedure consisted in taking several
exposures at the different programme wavelengths on each night and
then coadding them to produce a higher signal-to-noise final frame.
The observations were performed on the following emission lines of
\hh\ arising in the K band; v=1-0\,S(1) [2.1218\mic],
1-0\,S(0)[2.2235\mic], 2-1\,S(1) [2.2477\mic], 2-1\,S(2)[2.1540\mic] ,
2-1\,S(3)[2.0650\mic], 3-2\,S(1) [2.3846\mic], 3-2\,S(3) [2.2014\mic],
and the 4-3\,S(3) [2.3445\mic]. In addition, we have observed the
\brg\ transition at 2.166\mic\ and two emission lines of iron:
\irontwo\ and \ironsix\ arising on the J and H bands, respectively.
Table~1 provides a list of all the line observations together with the
observational parameters for each line transition.

One of the main goals of the project was to observe the
high-excitation \hh\ lines (with upper level energies above 13000 K)
of molecular hydrogen in the K band, especially the ones arising from
vibrational levels v$'$=3 (e.g. the 3-2\,S(3) [2.2014\mic]) and v$'$=4
(e.g. the 4-3\,S(3) [2.3445\mic]).  Making exposures at such low-light
levels that are characteristic of these faint lines in shocked regions
required a special technique since several effects such as the
non-linearity of the array at low-light levels may be comparable with
the photon noise level (Gredel and Moorwood 1991).  The method
employed consisted in taking several frames at the same wavelength but
slightly shifted by 1/2 and 1 pixel, thus ensuring that the same flux
is sampled by different pixels on the array.  This also helps to
prevent the spread of bad pixels to the final image. To produce the
final reduced image we align the frames to a common wavelength grid
and coadd up the exposures to gain sufficient signal-to-noise on the
emission line.

Three photometric standard stars were routinely observed throughout
the night at the same object wavelengths and airmass to allow a proper
cancellation of telluric absorption features and to flux calibrate the
object. The stars observed were: Beta Car (HR 3685), Eta Cma (HR 2827)
and Iota Car (HR 3699).  The relative variation of the standard star
flux over the three night period, provided an estimate of the
uncertainty on the calculated flux of $\sim 10$\%.  After inspecting
the reduced standard star frames, we noted that at two wavelengths,
the standard star continuum displayed deep absorption features at
2.166\mic\ and 1.642\mic.  These features are known to be associated
with the absorption by the stellar photosphere of the hydrogen
recombination radiation.  The lines in question are the \HI\ [n=7-4]
and \HI\ [n=12-4], for the 2.166\mic\ and 1.642\mic, respectively and
were artificially removed by linear interpolation between the line
edges, a common practice since the error committed is negligible
compared to the star's continuum absolute flux.

The flux calibration procedure required a flux table for each standard
containing a properly sampled representation (lambda vs.\ flux) of the
continuum for each standard star.  Such file was constructed for each
star based on the broad band flux given in the SIMBAD\footnote{The
  SIMBAD database is operated by the Centre de Donn\'{e}es
  astronomiques de Strasbourg, France} database.  Using this table, it
is possible to reproduce the star's flux distribution over the
wavelength region of interest and to use this curve to calibrate the
flux counts in the object frames.  All data were primarily reduced
with specific ESO routines available in {\sc midas} for IRSPEC
observations.  The later stages of data reduction that include
standard procedures (sky subtraction, flat-fielding, wavelength and
flux calibration) were assisted with routines from the {\sc starlink}
project.

With the current IRSPEC standard setup, a slight spatial displacement
of the objects occurs between frames taken at different wavelength
settings.  This is known to be caused by changing the grating settings
between the different exposures (Gredel and Moorwood 1991).  Thus, in
order to correct for this effect, we have determined the relative
spatial shift between the different wavelengths frames by calculating
the centroid position in the spatial direction for the observed star
continuum.  This was achieved by fitting a gaussian (instrumental
spatial profile) to the spatial profile of the star's image for all
wavelengths.  We found a shift of $+3.5$ pixels for the \irontwo\ 
image map and of $-2.5$ pixels for the \ironsix\ image map relatively
to the \hh\ 1-0 S(1) map.  All the other \hh\ lines have intermediate
shifts ranging from $-0.5$ at 2.065\mic\ to $+2.5$ at 2.3846\mic.  All
the calculated spatial shifts between the frames are shown in column 5
of Table~1.

\section{Results}

A total of eight \hh\ transitions were observed during the first two
nights.  These arise from vibrational levels: v=1, 2, 3 and 4, and
have all rotational jumps of $\Delta J=+2$.  Table 1 shows the
parameter list of the observed emission lines.  For each line
observed, this table shows in the fourth column the total exposure
time of the final reduced frame.  In addition, the fifth and sixth
columns give respectively, the spatial shift used to correct the
different grating settings to the same offset and the velocity
resolution for each frame.

The heliocentric velocity correction for the time the observations
were performed ($-3$\kms) is negligible when compared with the width
of a pixel (48--98\kms). Thus, we have chosen not to make any
corrections on the calculated position-velocity maps after converting
the wavelength scale to the radial velocity scale on the maps. This
conversion was applied only to the \hh\ 1-0 S(1) and the two \iron\ 
maps since only these are bright enough to compute the radial velocity
of emitting gas.  The velocities are thus measured with respect to the
Bok globule which is believed to have a heliocentric radial velocity
$\sim +22$\kms\ (Hartigan et al. 1993).

\subsection{The line emission}
\label{sec:hh}

Figure~2 shows the position-velocity contour emission map observed for
the \hh\ 1-0~S(1) transition at 2.1218\mic. This transition is one of
the strongest emission lines of \hh\ to arise in the K band and has
been extensively used to map regions of shocked gas in molecular
clouds (see for example Eisl\"offel 1997, for a recent review).  Some
unwanted features are also present in this emission map; some CCD
columns to the right and left of the main central object chain show
vertical and narrow linear emission structures that stand above the
background noise level.  These are probably due to the combined effect
of poor cancellation of sky lines (due to the intrinsic sky emission
line intensity changing over the integration period) in the reduction
procedure and to the slight movements of the grating after changing
(wavelength) position.  The wavelength position of these features were
compared with the list of OH airglow emission lines given in Olivia
and Origlia (1992) and can be positively identified with the residual
sky emission lines left over from the reduction procedure.  This
allows to check the on-line (mechanical) wavelength calibration
(usually precise to $\sim 1$ pixel).  Using the two brightest OH lines
in our spectra at 2.11710\mic\ [(9,7)~$R_1$(2)] and 2.12440\mic\ 
[(9,7)~$R_1$(1)], we have calculated the deviation of their profile
centroids to the ones listed in Oliva \& Origlia (1992).  This gives a
r.m.s. deviation error of 0.46 pixels and therefore imply an
associated uncertainty of $\approx 30$\kms\ for the measured radial
velocities.

Apart from the continuum emission that is associated with the HH46
reflection nebula near the infrared source located close to the middle
of the frame (rows $\sim$22--25), there are several objects clearly
seen as bright emission knots aligned vertically in the middle of the
array.  These objects are labeled in Figure~2 and their relative
positions can be identified using the maps of Eisl{\"o}ffel et~al.
(1994) and Micono et~al. (1998).

Figure~2 shows the northeastern blueshifted outflow to be mainly
composed of two bright emission knots, HH46 and HH47A. In the middle
of the frame, HH46 appears as a bright knot blended with the nebular
continuum emission from the ``reflection nebula'' centered at $\sim$
row 24. The emission from HH46 appears blueshifted with a heliocentric
radial velocity of about $-70$\kms, which is at the lower end of the
radial velocities measured for [SII] and H$\alpha$ emission (Dopita et
al.  1982, Meaburn \& Dyson 1987, Reipurth \& Heathcote 1991, Hartigan
et al. 1993, Morse et al. 1994) which lies in the range $\approx -80$
to $-170$\kms.

Further north, at the top of the PV map, there is a bright knot which
we associate with the infrared counterpart of the high-excitation
HH47A object. This object is commonly associated with the brightest
working surface of the northeastern outflow and is located at a
distance of about 65\arcsec\ northward of HH46.  The centroid velocity
of this bright emission peak is seen at a heliocentric radial velocity
of about $-100$\kms, which correlates well with the position of the
optical, H$\alpha$ and [SII], emission knots seen blueshifted at $\sim
-80$ to $-140$\kms\ (Dopita et al. 1982, Meaburn \& Dyson 1987,
Hartigan et al. 1993, Morse et al. 1994).  HH47A is considered the
best example of a bow shock/Mach disk typical interaction, even though
it is a secondary ejection from the source that is ploughing through a
pre-accelerated and dense medium -- the wake of an earlier shock now
observed as HH47D. Imaging the HH47A shock in the near-infrared \hh\ 
1-0 S(1) line shows that it splits up in two regions: a bright knot Z
and a fainter nebulosity to the south named as knot Y (Micono et~al.
1998).  In the region bridging HH46 to HH47A, there appears to be no
significant \hh\ emission in our maps, apart from the faint emission
filament located north of HH46 and extending up to a distance of
20\arcsec\ (rows $\sim 25$--$32$) at essentialy the same radial
velocity.  This is in good agreement with near-infrared maps; indeed,
only a fainter emission patch can be seen to the west of the knot Z
(Micono et~al. 1998), which lies outside our slit. Even though this
object lies at the edge of the frame, where the detector array may
suffer from vignetting, we have, nevertheless, measured the line
intensity and estimated the integrated column density for the
brightest \hh\ lines in this object (see Table~3).


The most striking feature of the \hh\ near-infrared maps, however, is
the bright and extended emission at the base of the redshifted outflow
located south of HH46.  This emission is associated with an extensive
counter-jet shock clearly visible to the southwest of the HH\,46/47
source in the \hh\ 1-0~S(1) image presented by Eisl{\"o}ffel et~al.
(1994). This highly collimated counter-jet is not detected in
H$\alpha$ emission and was only discovered by Reipurth and Heathcote
(1991) in their [SII] $\lambda\lambda$6716/6731\AA\ emission map as
being composed of three emission knots extending 11\arcsec\ on the
sky.  The counter-jet faint appearance in the optical is probably due
to the increased extinction at optical wavelengths and the fact that
the counter-flow is moving towards the interior of the Bok globule.
However, it is readily seen as the brightest object in our
near-infrared PV maps since the extinction decreases about an order of
magnitude from optical to infrared wavelenghts.  This counter-jet is
known to be at the base of a larger (infrared) bow-shock that extends
deeper into the interior of the Bok globule for about 2\arcmin\ until
it escapes out of the globule at HH47C (also seen in the infrared;
e.g. Eisl\"offel et al. 1994).  Figure~2 shows the counter-jet to be
resolved in to {\em two\/} spatially blended emission knots but that
appear clearly displaced in heliocentric radial velocity. The extent
of this counter-jet is 22\arcsec\ on the sky or 0.037 pc at the
distance of HH\,46/47 (using a distance to the globule of 350 pc,
Micono et al.  1998).  The peak-to-peak separation of these knots is
0.015 pc (8.8\arcsec\ on the sky) which correcting for projection
effects of the outflow orientation angle with respect to the plane of
the sky ($\sim$35 degrees, Eisl{\"o}ffel and Mundt 1994) gives a
separation distance of 0.026 pc.

The most interesting feature seen in the counter-jet is the apparent
radial velocity separation of the two knots. Using the nomenclature of
Micono et al. (1998), we identify the peak closer to the IR source as
``c-jet a'' which we will refer hereafter as the Counter-jet North,
and ``c-jet b+c+d`` as the jet formed by three knots (unresolved in
our images) which correspond to the southern knot and we will refer to
it as the Counter-jet South. The brightest emission peak, Counter-jet
North, appears at a distance of $\sim$ 15\arcsec\ from HH46 with the
emission centered at a heliocentric radial velocity of $\sim +27$\kms,
suggesting that the counter-jet starts moving near the systemic
velocity of the globule as it travels away from the source and
gradually increases its radial velocity by $\approx 28$\kms\ within
$\sim 9$\arcsec\ along the southwestern emission lobe where the second
peak, Counter-jet South, appears redshifted by $\sim +55$\kms.  We
note also that this same pattern for acceleration of the gas is also
seen in the proper motion data of Micono et al., where the proper
motion knot shift is of order 0.082 arcsec yr$^{-1}$ for knot c-jet a
(Counter-jet North) and 0.138 arcsec yr$^{-1}$ for knot c-jet d
(farthest part of Counter-jet South).

We have also detected some emission at the southern edge of the slit
(row 1) which we identify as ``knot~5'' in the nomenclature of
Eisl{\"o}ffel et~al. (1994, see Fig.~2a) or ``knot~g'' using the
latest near-infrared mapping of this region (Micono et al. 1998). We
note that the position of this knot in the PV map suggests that the
redshifted jet is flowing at essentially the same heliocentric radial
velocity as the Counter-jet South knot.

Although the counter-jet flow shows up very brightly in the \hh\ 
1-0~S(1) line emission map, it gets fainter with higher excitation
energy of the \hh\ levels (see Fig. 3). Note how the counter-jet is
just barely visible in the 2-1~S(2) emission map and completely
disappears on the noise in the 3-2~S(3) emission map.  The \hh\ 
2-1~S(1) contour emission map reveals that the counter-jet is still
well defined, showing the same characteristic ``double peak'' shape as
in the 1-0~S(1) map.  Some fainter emission is still visible at the
HH47A position but HH46 is now barely seen above the continuum
emission from the reflection nebula.  The \hh\ 1-0~S(0) emission map
is very similar to the 1-0~S(1) map with HH46 standing well out from
the nebular continuum.  This map also shows some residual OH lines
that could not be perfectly subtracted from the final reduced object
frame.  However, this does not affect the flux measurements in
Section~\ref{sec:fluxes} since the OH lines are non-coincident with
the \hh\ line.  The emission maps obtained for the 3-2~S(1) and
4-3~S(3) lines (not shown) show substantially fainter emission from
these objects implying a lack of high excitation \hh\ in the gas since
these lines arise predominantly from hot spots in the flow where the
temperature is high enough to populate significantly the upper energy
levels of the \hh\ molecules.

Given this, further analysis is thus restricted to the brightest four
lines, 1-0~S(1), 2-1~S(1), 1-0~S(0) and 3-2~S(3).  Table~3 shows the
physical parameters of these \hh\ lines used in this work.

In order to fully investigate the emission spectra from the
counter-jet we coadded all the detector rows comprising the entire
counter-jet flow (rows 10--20) and fitted simultaneously two gaussians
to the resulting spectra.  Thus, the reference to Counter-jet North,
in column 1 of Table~4, refers to the gaussian solution to the fit of
the emission peak closer to the source whereas the reference to
Counter-jet South refers to the gaussian solution to the fit of the
emission peak far from the source.  The reference to the Counter-jet
Total was obtained by fitting a single gaussian profile to the spectra
but allowing a broader FWHM to account for the line blending.  The
computed three-fold \hh\ line fluxes from the counter-jet are
presented in Table~4.


We have also searched for Brackett-$\gamma$ emission from the
counter-jet, but nothing was found. The 2.166$\mic$ map only showed
very weak continuum emission near the position of HH46 suggesting that
the observed infrared emission is produced mainly by dust scattering
in the diffuse reflection nebulae associated with HH46.

\subsection{\hh\ excitation temperature}
\label{sec:fluxes}

The physical conditions for the excitation of \hh\ emission can be
quantified by calculating the \hh\ level populations that are excited
in the gas.  Since \hh\ is a homonuclear molecule it may only be
excited through electric quadrupole radiation. In these conditions it
is usual to assume the gas to be optically thin at these wavelengths,
greatly simplifying the radiation transfer.  Thus, the line intensity,
$I_{v,J}$, for an \hh\ transition out of level ($v,J$), is directly
proportional to the transition energy, $h\nu_{v,J}$, the Einstein
coefficient, $A_{v,J}$ and the column density $N_{v,J}$ of molecules
excited in that particular level,

\[ I_{v,J} = h\nu_{v,J} A_{v,J} N_{v,J} / 4\pi \]

where h is Planck's constant and $\nu$ is the transition frequency.
The values for $A_{v,J}$ are those from Turner et al. (1977) and the
values for $\nu_{v,J}$ are taken from Dabrowski (1984).  For a
Boltzmann distribution of the vibration-rotation levels (thermal
distribution) at some local excitation temperature $T_{\rm{ex}}$, the
population density, $N_{v,J}$, of level ($v,J$), is proportional to
the statistical weight (level degeneracy) $g_{v,J}$ and the Boltzmann
factor.  Thus,

\[ N_{v,J} = g_{v,J}~ e^{-\frac{E_{v,J}}{kT_{\rm{ex}}}} \]

where $E_{v,J}$ is the transition energy, $k$ is Boltzmann's constant
and $T_{\rm{ex}}$ is the excitation temperature.  The excitation
temperature can now be estimated from the data by plotting
$N_{v,J}/g_{v,J}$ {\it versus} the upper level energy of the
transition $E_{v,J}$.  The coefficient of the exponential fit to this
column density excitation diagram is proportional to
$T_{\rm{ex}}^{-1}$.

Figure~4 shows the \hh\ excitation diagrams calculated for the
counter-jet emission using the data in Table~4.  Since the counter-jet
appears spatially extended and constituted by two emission peaks with
different velocity components, we have plotted in Fig.~4 three
separate data sets corresponding to the total flux emission from the
whole counter-jet (Counter-jet Total), the northern peak (Counter-jet
North) and southern peak (Counter-jet South) emission.  The excitation
temperature, $T_{2,1}$, indicated by the 1-0~S(1) to the 2-1~S(1) line
ratio in the Counter-jet North position is 2100 K while in the
Counter-jet Total position it is 1800 K (see Table~4). A lower value
of 1300 K is found for the Counter-jet South position.  In Figure 4,
the solid line represents a least square fit through all the data
points, providing an excitation temperature of $T_{\rm{ex}} = 2600$ K
for the \hh\ gas at both the Counter-jet North and Total positions.
The typical error of these estimates is given by the least squares fit
and is of the order of 300 K.

Also shown in Table~4 are the \hh\ line intensities and corresponding
column densities for the HH46 and the HH47A objects.  Unfortunately,
at these positions the low signal-to-noise obtained in the higher
excitation transitions ($v \ga 3$) hampers a more accurate
determination of the excitation temperature. The excitation
temperature for these objects are $T_{2,1}$=1200~K and
$T_{2,1}$=1800~K for the HH46 and HH47A objects, respectively.

\subsection{The \iron\ emission}

The forbidden emission from \iron\ at 1.257\mic\ ($ a^4 D_{7/2} - a^6
D_{9/2} $) and 1.644\mic\ ($ a^4 D_{7/2} - a^4 F_{9/2} $) was observed
on the third night at the same slit position used for the \hh\ line
observations to seek higher excitation regions within the flow and
also to provide an estimate of the foreground extinction in the
emitting gas.  The position-velocity emission maps obtained for these
two transitions are shown in Figure~5.  Superposing both maps on a
constant velocity grid makes all the bullet features coincide to
within half pixel which gives an indirect way of checking the spatial
shift correction performed on these maps. Since these \iron\ PV maps
show important morphological and kinematical differences from the \hh\ 
maps presented previously, we will next examine them in detail.


In the northeastern blueshifted outflow region, we see from Fig.~5,
that this part of the flow appears composed of four bright \iron\ 
knots with V$_{Hel} \sim -250$ to $-280$\kms, from which the northern
most is clearly (spatially) associated with the HH47A object (at the
top edge of the array in the \ironsix\ map due to the spatial shift
occurred when changing grating positions).  This confirms the
detection made by Eisl{\"o}ffel et~al. (1994) of a strong
near-infrared \ironsix\ emission peak at the position of the optical
HH47A object.  We note that the radial velocity of the emitting \iron\ 
in the HH47A object ($-250$\kms) is much greater than the
corresponding \hh\ emission at $-100$\kms.  On the other end, and at
the base of this blueshifted outflow, the HH46 object shows up
brightly against a very weak continuum.  However, the most remarkable
features in this Figure, are the appearance of {\sl two\/} emission
\iron\ knots in the northeastern part of the HH\,46/47 jet, in the
region bridging HH46 to HH47A, contrasting sharply with the morphology
presented by the \hh\ emission maps (Figs.~2--3).  These two knots
appear closer to HH46 and their locations in the outflow seem to have
some parallel with the optical emission. Several high-resolution
optical images of this part of the outflow have been published that
show detailed structure across the whole region, from the base of the
outflow at HH46, passing through the bright HH47A knot and ending in
the HH47D object (Reipurth and Heatcote 1991, Hartigan et al. 1993,
Eisl{\"o}ffel and Mundt 1994, Heathcote et al. 1996).  The \iron\ maps
presented here show the existence of two relatively bright knots in
this part of the outflow which we can tentatively identify with the
optical knot B8 located at a distance of about 18\arcsec\ from HH46
(Eisl{\"o}ffel and Mundt 1994), and knot B0 (HH47B), located at a
distance of about 32\arcsec\ from HH46 (see Figure~6).  This is in
contrast with the faint \hh\ outflow which appears, in this region, to
fade away at approximately the position of the B8 \iron\ knot.  Thus,
it appears that the \iron\ emission is tracing a higher velocity
outflowing jet component with no relation to the lower velocity \hh\ 
jet.  The whole northern \iron\ outflow appears blueshifted at
heliocentric radial velocities between $V_{Hel} \sim -280$\kms\ (HH46,
B8 and HH47B) and $V_{Hel} \sim -250$\kms\ (HH47A).

The \ironsix\ emission was first observed in this object by
Eisl{\"o}ffel et~al.  (1994, see Fig.~2a). Their emission map,
however, shows a slightly different spatial distribution of \ironsix\ 
emission, especially regarding the HH47B object. While the blueshifted
knots associated with HH46 and B8 coincide in both maps, knot HH47B
was not detected in their contour map. Two reasons might explain this
fact. Either our map is slightly more sensitive or the evolution of
the gas emission over time may have determined the fading of the
emission knot. Either way, we note that both emission knots are
clearly visible in our \ironsix\ and \irontwo\ maps and are both above
(2$\sigma$) the noisy sky background.  Overall, the spatial morphology
of our \ironsix\ PV map closely matches the \ironsix\ image map of
Eisl{\"o}ffel et~al.\ although, the emission in our image is clearly
more extended and brighter along the main blueshifted outflow axis.


Turning to the southwestern region, there is a redshifted bright knot
which is clearly associated with the emission from the counter-jet.
As seen earlier, this counter-jet emission is elongated and marginally
resolved in to {\em two\/} \hh\ knots. However, we now see the \iron\ 
emission from the counter-jet as a {\em single\/} redshifted knot at
$V_{Hel} = +100$\kms.  Moreover, when we overlay the \hh\ and \iron\ 
emission (see Figures 6 and 7), we conclude that this \iron\ knot is
associated with the Counter-jet North peak and that no \iron\ emission
is seen from the Counter-jet South peak.  Notice also how this \iron\ 
knot seems to appear marginally closer to the outflow source than to
the bright Counter-jet North \hh\ knot (c.f Figure~7) while it appears
to be travelling substantially faster than the corresponding \hh\ knot.

\subsubsection{Reddening}
\label{sec:reddening}

The extinction can be calculated using the relative flux observed from
the \iron\ transitions since these two lines arise from a common upper
level of the Fe$^+$ ion ($a^4 D_{7/2}$) and the line ratio is
therefore independent of the number of excited ions on that particular
level state.  Thus, their intensity line ratio, $I_{1.257}/I_{1.644}$,
should depend only on the intrinsic properties of the Fe$^+$ ion,
provided that the lines are optically thin.  This requirement should
be true since the critical density for the \ironsix\ transition is
$n_{crit} \sim 3 \times 10^4$\cc\ (Oliva, Moorhood and Danzinger
1990).  The differential reddening between the two \iron\ lines,
E$_{J-H}$, can be computed from:

$$
I_{1.257}/I_{1.644} = \left(I_{1.257}/I_{1.644}\right)_0 \times
10^{-E_{J-H}/2.5}
$$

where $\left(I_{1.257}/I_{1.644}\right)_0$ is equal to 1.36; the
theoretical line ratio given by the Einstein spontaneous probabilities
of Nussbaumer \& Storey (1988).  The visual extinction is obtained
using $E_{B-V} \approx 3 \times E_{J-H}$ (Draine 1989) and assuming a
normal reddening law: $A_V = 3.1 \times E_{B-V}$.  The derived \iron\ 
line flux and extinction estimates are shown in Table~2 for the
observed emission peaks.  This table shows that the \iron\ bullet
associated with the Counter-jet North has a visual extinction of 7.05
\p\ 2.26 whereas an extinction value of 9.38 \p\ 1.49 is measured in
the bullet associated with HH46, closer to the obscured YSO. Given the
errors, these values indicate that the extinction is uniform
throughout the region.
  
Note also that we measure $A_V \sim 0$ for the northern \iron\ bullets
B8 and HH47B; given the uncertainties associated with the low flux
measured for these bullets.  This strongly suggests that the visual
extinction is negligible in this (blueshifted) northwestern part of
the outflow.  However, as noted by Oliva and Origlia (1992), the
\ironsix\ transition sits close ($\sim 40$\kms) to the relatively
bright OH (5,3) $R_1 (2)$ transition which hampers a more accurate
determination of the reddening.

\section{Discussion}

The intensity ratio of the $v$=1-0 S(1) line to the $v$=2-1 S(1) line
has been widely used to discriminate between two common types of
physical mechanisms for the excitation of HH objects: thermal (shocks)
and non-thermal (e.g. fluorescence by Lyman-$\alpha$ or UV continuum
pumping, \hh\ reformation).  HH objects are preferentially shock
excited but fluorescent excitation by Lyman-$\alpha$ pumping may also
be present (HH43; B{\"o}hm, Scott and Solf 1991, HH47A; Curiel et al.
1995, HH7; Fernandes and Brand 1995).  The 1-0/2-1~S(1) ratio derived
from the observations (Table~5) for the Counter-jet Total, Counter-jet
North and HH47A objects are $15.7 \pm 2.3$, $10.1 \pm 3.5$ and $16.4
\pm 2.5$, respectively. These values are consistent with shock
excitation (in which the 1-0/2-1~S(1) ratio $\ga$ 5) as fluorescent
excitation requires the 1-0/2-1~S(1) ratio to be $\la$ 2 in low
density regions. In high density regions the fluorescence 1-0/2-1~S(1)
ratio may be $\gg$ 2. However, this is unlikely to be the case here,
since these high density models predict large column densities in the
v=3 and v=4 levels and thus strong emission from lines such as the
3-2~S(3) and 4-3~S(3) which could not be observed from the data.
Under these conditions, the molecular hydrogen emission has its origin
behind dense shocks and the \hh\ excitation diagrams calculated for
the HH objects (Figure~4) are consistent with thermalised gas at a
temperature of $T_{2,1} \approx$ 1300--2100 K for the counter-jet
whereas HH46 and HH47A display $T_{2,1} \approx$ 1200 and 1800 K,
respectively. We note here that non-LTE, low density gas, may also
produce line ratios similar to those observed here, as we will discuss
below.

The densities required to thermalise the \hh\ levels up to 20\,000~K
are of order $10^6$--$10^7$\cc\ in a pure molecular gas (Burton et al.
1989).  These high densities can be attained in regions close to the
source from where the counter-jet emerges or in the dense relaxation
medium behind the shock front where the gas cools via atomic, ionic
and molecular line emission. Both pure hydrodynamic shocks (J-type)
and magneto-hydrodynamic shocks (MHD C-type) can produce significant
emission from \hh, although only fast (dissociative) J-type shocks can
produce observable \iron\ emission since C-shocks attain a lower
post-shock temperature ($\la 2000$ K) than J-shocks ($\gg 2000$ K).

Comparing the 1-0/2-1~S(1) line ratios with the shock models of Smith
(1995) and Eisloffel et al. (1996) yields a useful constraint on the
various critical shock parameters such as the pre-shock density and
the type and geometry of the shock. First, note from Table~5 that the
observed range for the line ratios of 10--17 (except for the noisier
Counter-jet South and HH46 objects) is consistent with a slow planar
J-shock with speed 8--9\kms. However, these line ratios can be also
reproduced by fast planar C-shocks with speeds of 35--40\kms. The
modelled 1-0~S(1)/3-2~S(3) line ratio of 447.7 for a 35 \kms\ C-shock
is nonetheless much greater than the observed value of only 45--50 and
thus planar C-shocks can be ruled out. On the other hand, the slow
J-shocks are much weaker than those required to excite the observed
\iron\ lines as we will see below.  On these grounds, we conclude that
neither {\em planar\/} type of shock can explain the observations,
except, perhaps, for the Counter-jet South and HH46 objects from which
no conclusions can be drawn due to the large errors in the observed
line ratios.

Constant temperature models were also compared with the observations.
The single (1T) and mixed temperature (2T) slab of gas models for both
LTE (n=10$^8$\cc) and non-LTE conditions (n=10$^4$\cc) yield specific
line ratios which are given in Table~5 (Smith, private communication).
The three chosen 1T models in Table~5 fail to account for both line
ratios regardless of the LTE conditions. Moreover, the excitation
temperature derived from the 1-0/2-1~S(1) line ratio ranges
1800--2100~K, whereas the least square fit to all the data points
implies a higher excitation temperature of 2600~K. Given this, we may
expect better results from the dual temperature models. In fact, an
inspection of Table~5, clearly indicates a closer match to the data.
However, there are strong physical constraints on these approach since
we would need a heating mechanism capable of maintaining the gas at
fixed temperatures of order of a few thousand Kelvin for a timescale
of years. At present, there is no valid explanation for such
mechanism, and thus we also rule out this possibility.

In agreement with the observed morphology for these type of objects,
we thus turn our attention to the excitation of both \hh\ and \iron\ 
emission in bow-shock models.  Strong \iron\ emission is often
interpreted as indicative of shock excitation (e.g Moorwood \& Oliva
1988) and can be expected in, for example, supernova remnants (McKee,
Chernoff \& Hollenbach 1984).  In bow-shocks, however, the high
excitation \iron\ emission derives from the bow-head, where the shock
is strongest (see e.g. HH1; Davis, Eisloffel \& Ray 1994). Further
downstream, in the lower excitation oblique bow wings, emission from
\hh\ can be excited in the softer (non-ionising) shocks. The presence
of \ironsix\ in the flow is thus considered a diagnostic of fast
dissociative shocks at the tip of the bullet that heads a post-shocked
and accelerated medium observed as a trailing \hh\ wake (see e.g.
Tedds, Brand and Burton 1998).

The bow shock emission structure can be modelled by a surface of
revolution which can be simply written as $z \propto r^s$ in
cylindrical coordinates, where $s$ is the free parameter that adjusts
the bow aperture (see Fig.~6 of Eisloffel et al. 1996).  This
morphology combines in a single structure, two distinct regions,
depending on the local physical conditions, especially the ion
fractions and the strength of the magnetic field. Thus, we need to
distinguish between J-type or C-type bow shocks.

The former, produces emission which is everywhere excited by pure
hydrodynamic type shocks, from the apex of the bow, where strong
J-shocks destroy the \hh\ molecules and ionises Fe, to the extended
\hh\ emitting wings downstream along the surface. The predictions from
these bows depend critically on the shape parameter $s$. The best
parameter fit models are shown in Table~5. In general, both LTE and
non-LTE models come close to adjust the observed line ratios. The best
fit bow shape parameter in these models range from 1.2 to 2.0
(paraboloidal) which is consistent with strong [FeII] emission over
much of the outflow.  Narrow line profiles are expected from these
type of shocks and can be used to discriminate between shock types,
since the larger shock front of C-type shocks will produce wider line
profiles. This, however, requires high-resolution spectroscopy data
and therefore with the present set of data we cannot rule out the
J-shock bow model for the Counter-jet or HH47A.

On the other hand, the C-type bow shock model differs from the above
picture only on the bow wings, where the softer C-type shocks produces
emission from \hh\ and CO.  The Orion bullets are the best studied
example of this structure, where a high-velocity leading \iron\ bullet
is ploughing through the medium leaving behind a wake of
lower-excitation \hh\ emitting gas generated at the warmer and oblique
C-type shocks at the bow flanks (Tedds, Brand and Burton 1999).  The
C-type bow-shock model fits shown in Table~5 can also be made to agree
with the present line ratios, but for a larger $s$ bow shape parameter
($1.5 < s < 2.0$). In this case, the bows are closer to paraboloids
and less pointed than the J-type bows.

The \ironsix\ surface brightness predicted for a fast (v $>$ 30\kms)
J-shock can be large due to the large column of warm gas generated by
UV absorption and \hh\ reformation in the post-shock layer (Smith
1994).  For a 100\kms\ shock with a pre-shock density of 10$^4$\cc, a
surface brightness of $1.5 \times 10^{-4}$\ergs\ for the \ironsix\ 
line can be achieved. This value is consistent within a factor of 3
with our measured value at the Counter-jet North peak of $4.6 \times
10^{-4}$\ergs. Note that this is as strong as the observed \hh\ 1-0
S(1) surface brightness of $4.4 \times 10^{-4}$\ergs. Such high
values, where the \ironsix\ /\hh\ 1-0S(1) line ratio are of equal
strength, have already been observed in other HH objects by
Stapelfeldt et al. (1991) and can be well modelled with fast J-shock
models (Smith 1994).

A non-dissociative J-shock (v $\la$ 22\kms), on then other hand,
produces smaller columns of warmer gas. We may estimate an upper limit
for the \ironsix\ line strength from the model employed by Smith
(1994) under the assumptions that the electron fraction is $\sim
10^{-3}$ and the abundance of Fe$^+$ is $\sim 10^{-6}$. For the
fastest non-dissociative shock (with v = 22\kms), the line strength
predicted is $< 4.2 \times 10^{-5}$\ergs, a value which is about 10
times smaller than observed.  For a 15\kms\ shock speed, the \ironsix\ 
intensity drops down by a factor of 2 and therefore we rule out the
slow J-shock model for the generation of excited \iron.

Downes (1997) has modelled the \hh\ and [SII] $\lambda$6731 emission
from the HH\,46/47 counter-jet using an hydrodynamical code to predict
intensity maps and line profiles. The simulated emission showed the
appearance of numerous mini-bowshocks throughout the outflowing jet
resulting from Rayleigh-Taylor instabilities.  Moreover, the
synthesised [SII]$\lambda$6731 and \hh\ 1-0~S(1) line profiles derived
from their emission maps of the counter-jet, yield clear asymmetries
in the profiles with the near-infrared line peaking at 3--4\kms, while
the optical line peaks at about 150\kms.  Since the \hh\ emission
appears only at the bow flanks, it displays a low velocity component;
in our data +27 and +55\kms, for the Counter-jet North and South
objects, respectively.  In contrast, the \iron\ knot coincident with
the Counter-jet North object, is seen peaking at a higher velocity of
+100\kms. This agrees well with the bow shock scenario, where the high
excitation lines of [FeII], [SII] and H$\alpha$, are produced in the
strong and high speed shocks near the bow apex, yielding higher normal
gas velocities of $\sim 100$--$150$\kms.

Looking at the relative position of the \iron\ and \hh\ emission in
HH47A (see Figs. 6 and 7), we see the \iron\ peak ahead about
5\arcsec\ from the \hh\ peak. The spatial calibration is accurate to
less than half pixel ($\sim 1$\arcsec) and thus we consider the shift
to be genuine. This translates to a separation distance of $\sim 3
\times 10^{16}$ cm from the \iron\ emitting section to the \hh\ 
emitting section. This separation is characteristic in bow-shocks
viewed at orientation angles $> 30$ degrees to the line of sight
(Smith 1991) and thus we suggest that HH47A can be well modelled by a
bow shock working surface, in a similar fashion to HH99B (Davis, Smith
\& Eisloffel 1999). Note, however, the large radial velocities
achieved for the \iron\ ($-250$\kms), H$\alpha$ ($-130$\kms) and even
\hh\ ($-100$\kms), whose molecules should be all dissociated at these
high (v $\ga$ 50\kms) speeds. To produce \hh\ emission at these
extreme velocities, it is necessary that the pre-shock medium had
already been set in motion by some previous event (e.g. through the
passage of an earlier shock wave which may be accounted by the faint
HH47D object), so that the relative velocity between the two flows
stays lower than the dissociative limit of the \hh\ molecule: 25\kms,
for a J-shock or 50\kms, for a C-shock (Smith 1991).  Unfortunately in
this work, we have not reached the required sensitivity in the 3-2
S(3) line in HH47A to help discriminate between these two types of
shocks spectroscopically.  Thus, on geometrical grounds, we expect
HH47A to be a classical bow shaped shock, with a dissociative J-shock
section at the bow apex (which produces the \iron\ and the optical
emission), plus a longer C-shock wing section which produces the lower
excitation \hh\ emission.  With the measured tangential velocities in
this object of the order $\sim$ 180\kms\ (both optical and
near-infrared), a bow shock velocity of $\approx 220$\kms\ can be
inferred.

The emission from HH46 is partially blended with the reflection nebula
which causes the \hh\ emission to be scattered and thus appear weaker
in contrast to the brighter \iron\ emission.  The 2-1~S(1) line
emission from HH46 is somewhat below our expectations since the
1-0~S(1) remains strong. In fact, the 2-1/1-0 S(1) line ratio is only
0.014 compared with, for example, 0.061 in HH47A.  We suggest that we
may be observing only the emission tail of HH46, which produces weaker
shocks, and thus explaining the weaker \hh\ 2-1~S(1) emission observed
and also the $-70$\kms heliocentric velocity shift observed for this
object.  Moreover, the \iron\ emission peaks at $-280$\kms, implying
very fast bow shock speeds near the source. In fact, a shock which
produces \iron\ efficiently, will likely destroy the \hh. For \hh\ to
survive the extreme shock velocities observed in HH46, we therefore
require a mechanism that can soften the shock. Commonly cited in the
literature are mechanisms such as the magnetic precursor model, the
entrainment of material from the wind/ambient material interface or
the high Alfven Mach number shock absorbers (bow shocks with a high
magnetic field). Distinguishing between these physical processes is
beyond the scope of this paper with the available set of data.
Obtaining higher resolution data will certainly help to establish the
importance of these mechanisms.

\section{Conclusions}

The physical conditions along the redshifted counter-jet and the
blueshifted HH46 to 47A outflow were investigated through observation
of velocity-position diagrams and spectra in the J (\ironsix\ and
\irontwo) and K (several \hh\ transitions) infrared bands.

The observed \hh\ line emission confirms the existence of a bright and
extended redshifted counter-jet outflow southwest of HH46.  Moreover,
there seems to be compelling evidence for acceleration of the
counter-jet from 27\kms\ to 55\kms\ along the southwestern emission
lobe as it travels away from the source.

The observed \iron\ emission suggests a near uniform extinction in the
HH46 and the counter-jet region of $\approx 7$--9 visual magnitudes,
whereas the extinction calculated for the blueshifted knot B8 and
HH47B objects is negligible.

The \hh\ excitation temperatures derived from the 1-0/2-1~S(1) ratio
are in the range 1800--2100~K for all the objects, with the exception
of the HH46 object for which the excitation temperature is 1200~K. The
unusual low 2-1~S(1) line flux in this object was probably caused by
the limited region covered by our slit in this object which may have
sampled only the wing sections of the bow-shock structure.  The
inclusion of the 3-2~S(3) in the \hh\ excitation diagrams for the
counter-jet shock yields an excitation temperature for the gas of
$T_{\rm{ex}} \approx 2600$K.

The non-detection of higher energy transitions of \hh\ in the K band,
precludes several non-thermal excitation mechanisms such as
fluorescence or \hh\ reformation. The observed 2-1/1-0~S(1) and
3-2/1-0~S(1) line ratios were found consistent with models of J and
C-type bow shocks, whilst ruling out planar shocks.  In the
counter-jet, the shape parameter for C-type bows can be constrained to
the range $1.5 < s < 2.0$ whereas J-type bows need to be made
extremely wide, $1.2 < s < 1.5$, to adjust well our data.

\section*{Acknowledgements}
This work was supported by the Portuguese Funda\c c\~ao para a Ci\^encia
e Tecnologia (grant PESO/P/PESO/1196/97).
I thank Chris Davis and Michael Smith for some helpful suggestions 
to improve this manuscript.

\bsp

\end{document}